\documentclass[iop]{emulateapj}

\shorttitle{The History of a Quiet-Sun Magnetic Element Revealed by IMaX/\textsc{Sunrise}}
\shortauthors{Requerey et al.}

\begin{document}


\title{The History of a Quiet-Sun Magnetic Element Revealed by IMaX/\textsc{Sunrise}}


\author{Iker S. Requerey, Jose Carlos Del Toro Iniesta, and Luis R. Bellot Rubio}
\affil{Instituto de Astrof\'{i}sica de Andaluc\'{i}a (CSIC), Apdo. de Correos 3004,
    18080 Granada, Spain}
\email{iker@iaa.es}
\author{Jos\'{e} A. Bonet}
\affil{Instituto de Astrof\'{i}sica de Canarias, Avda. V\'{i}a L\'{a}ctea s/n, La Laguna, Spain}
\affil{Departamento de Astrof\'{i}sica, Universidad de La Laguna, E-38205 La Laguna, Tenerife, Spain}  
\author{Valent\'{i}n Mart\'{i}nez Pillet}
\affil{National Solar Observatory, Sacramento Peak, Sunspot, NM 88349, USA.}
\affil{Instituto de Astrof\'{i}sica de Canarias, Avda. V\'{i}a L\'{a}ctea s/n, La Laguna, Spain}
\author{Sami K. Solanki}
\affil{Max-Planck Institut f\"{u}r Sonnensystemforschung, 37191, Katlenburg-Lindau, Germany}
\affil{School of Space Research, Kyung Hee University, Yongin, 446-701 Gyeonggi, Korea}
\author{Wolfgang Schmidt}
\affil{Kiepenheuer-Institut f\"{u}r Sonnenphysik, Sch\"{o}neckstr. 6, D-79104, Freiburg, Germany}



\begin{abstract}

Isolated flux tubes are considered to be fundamental magnetic building blocks of the solar photosphere. Their formation is usually attributed to the concentration of magnetic field to kG strengths by the convective collapse mechanism. However, the small size of the magnetic elements in quiet-Sun areas has prevented this scenario from being studied in fully resolved structures. Here we report on the formation and subsequent evolution of one such photospheric magnetic flux tube, observed in the quiet Sun with unprecedented spatial resolution (0\farcs 15 - 0\farcs 18) and high temporal cadence (33 s). The observations were acquired by the Imaging Magnetograph Experiment (IMaX) aboard the \textsc{Sunrise} balloon-borne solar observatory. The equipartition field strength magnetic element is the result of  the merging of several same polarity magnetic flux patches, including a footpoint of a previously emerged loop. The magnetic structure is then further intensified to kG field strengths by convective collapse. The fine structure found within the flux concentration reveals that the scenario is more complex than can be described by a thin flux tube model with bright points and downflow plumes being established near the edges of the kG magnetic feature. We also observe a daisy-like alignment of surrounding granules and a long-lived inflow towards the magnetic feature. After a subsequent weakening process, the field is again intensified to kG strengths. The area of the magnetic feature is seen to change in anti-phase with the field strength, while the brightness of the bright points and the speed of the downflows varies in phase. We also find a relation between the brightness of the bright point and the presence of upflows within it.


\end{abstract}


\keywords{Convection --- Sun: photosphere --- Sun: granulation --- magnetic fields --- techniques: polarimetric --- spectroscopic}



\section{Introduction}

The interaction between convection, radiation, and magnetic field in the electrically conducting solar plasma leads to the creation of a rich variety of magnetic structures. Many of these have kG field strengths and range in size from the largest sunspots, tens of Mm in size, down to the smallest network and internetwork structures, i.e., ``magnetic elements'' on spatial scales of 100 km or less. The observations of decaying sunspots into smaller structures as well as the formation of pores from the accumulation of smaller magnetic features, has led to the notion that magnetic elements are fundamental entities of magnetic flux from which larger structures are assembled \citep[see, e.g.,][for reviews]{1993SSRv...63....1S,2009SSRv..144..275D}. 

The formation of magnetic elements is thought to be well understood from a theoretical point of view. It is generally accepted that the first step in producing such flux tubes is the \textit{flux expulsion} mechanism. As suggested by \citet{1963ApJ...138..552P} and \citet{1964MNRAS.128..225W, 1966RSPSA.293..310W}, the magnetic flux is advected by horizontal flows and concentrated in convective downflow areas, roughly up to the equipartition field strength (300-500 G), for which the magnetic energy density equals the kinetic energy density of the gas flow. These equipartition flux concentrations reduce the convective heat transfer, leading to super adiabatic cooling \citep{1978SoPh...59..249W, 1979SoPh...62...15S}.  This evacuates the flux tube, in such a way that the gas pressure of the surrounding plasma compresses the flux concentration until kG field strengths are reached \citep{1978ApJ...221..368P, 1979SoPh...61..363S}. This process is known as \textit{convective collapse}, which is thought to be the fundamental step of flux-tube creation.

Later, numerical studies revealed further details of the final state of the magnetic feature. \citet{1985A&A...143...39H} found that non-adiabatic effects arising from the radiative exchange between the flux tube and the external medium lead to overstable oscillations as the final state of a collapsed flux tube.  On the other hand,   \citet{1999ApJ...522..518T} reached a static solution, and showed that if the downflow in a collapsing flux tube becomes strong enough, an upward-traveling shock can develop as the downward flow bounces back in the dense deeper layers. This ``rebound shock'' reverses the magnetic flux intensification, and may lead to the dissolution of the magnetic flux concentration.

All these results are based on one-dimensional calculations, and rely on the thin flux-tube approximation. \citet{1998A&A...337..928G} made use of two-dimensional numerical simulations to study the interaction between the surrounding convective flow and the flux tube. They were the first to find the rebound shock solution for initial field strengths of 400 G. However, this result changes when the initial field is weaker, for which the flux sheet reaches a stable state rather than being dispersed. \citet{1999ASPC..184...38S} used similar numerical simulations and found a more ``quiescent phase'' during the time period between the formation of a magnetic flux tube and its dissolution or reformation. During this phase the magnetic field strength remains quite constant and the flux tube exhibits small internal gas motions. As a consequence of the interaction with the surrounding granular convection, the flux concentration moves laterally, bending and swaying, gets ``squeezed'', and during most of the time is bordered by strong, narrow downflows. These strong downflows get narrower with depth and accelerate strongly until they evolve into ``jets'' \citep{1998ApJ...495..468S}. They are maintained by the cooling of the gas surrounding the flux concentration through radiative heat losses into the magnetic structure \citep{1984A&A...139..435D}. 
More recently, \citet{2011ApJ...730L..24K} carried out two-dimensional radiation MHD simulations, and showed that these downflow jets can indeed excite a downflow within the magnetic flux concentration, which rebounds and develops into an upward-travelling shock front. Through this mechanism, the atmosphere within the tube oscillates at the acoustic cutoff frequency. Furthermore, \citet{2012ApJ...746..183J} found that upwardly propagating acoustic waves are ubiquitous in quiet-Sun magnetic bright points and 3-D MuRAM \citep{2005A&A...429..335V} simulations.

From an observational point of view, spectropolarimetric evidence of convective collapse and subsequent destruction of magnetic flux by an upward-moving front in the quiet Sun was reported by \citet{2001ApJ...560.1010B}. Magnetic flux intensification events have also been observed with the Hinode spectropolarimeter. First, a single event by \citet{2008ApJ...677L.145N}, where a strong downflow is detected while field strength intensifies and a bright point appears followed by a transient upflow, and then, a statistical analysis of 49 convective collapse events by \citet{2009A&A...504..583F}. 

The interaction between magnetic fields and convection is important to understand the formation and evolution of magnetic structures on the solar surface. \citet{1989SoPh..119..229M} observed that the presence of isolated Network Bright Points (NBPs) disturbs the surrounding granules which elongate in the direction of the magnetic features, forming a characteristic ``daisy-like'' structure. This granular pattern is formed as the small bright point appears while the surrounding granules converge \citep{1992SoPh..141...27M}. \citet{1997ApJ...478L..45B, 2000ApJ...535..489B} found from the inversion of full Stokes profiles of the Fe \textsc{i} 630 nm lines that magnetic flux tubes in facular regions are surrounded by intense downdrafts, and suggested that these downdrafts produce downflows of lesser magnitude in the tube interior. Close to small magnetic flux concentrations, \citet{2004ApJ...604..906R} observed strong, narrow ($<$0\farcs 2) downflow plumes at the edge of many small flux tubes, while there was little gas motion inside the flux concentration, confirming earlier results showing almost unshifted Stokes V zero-crossing in network and plage regions \citep{1986A&A...168..311S,1997ApJ...474..810M}. 

Isolated magnetic elements are the key to understanding a variety of solar structures, like plages, or the network. Unfortunately, these basic units are generally so small that they have mainly been studied using indirect techniques, either through the interpretation of Stokes spectra of the unresolved feature, or using their association to G-band bright points. Recently, the Imaging Magnetograph eXperiment \citep[IMaX;][]{2011SoPh..268...57M} launched on board the \textsc{Sunrise} balloon-borne solar observatory \citep{2010ApJ...723L.127S,2011SoPh..268....1B,2011SoPh..268..103B} allowed photospheric quiet-Sun magnetic flux tubes to be spatially resolved even in the quiet Sun \citep{2010ApJ...723L.164L, 2012ApJ...758L..40M}. 

Here, we take advantage of these unprecedented high quality observations to report on the formation and evolution of a small kG flux concentration and its interaction with the surrounding granulation. The data suggest that the magnetic element is formed by advective coalescence of small-scale flux patches and a subsequent convective collapse phase. Once formed, the evolution of the mature flux tube is much more complicated than that explained by static flux-tube models. Many different phenomena are involved, namely: converging granules and granular fragments, downflow jets, bright points (BPs), oscillations in all basic physical quantities, small-scale upflow plumes, etc.

\section{Observations and data analysis}


\begin{figure}
\includegraphics[scale=.45]{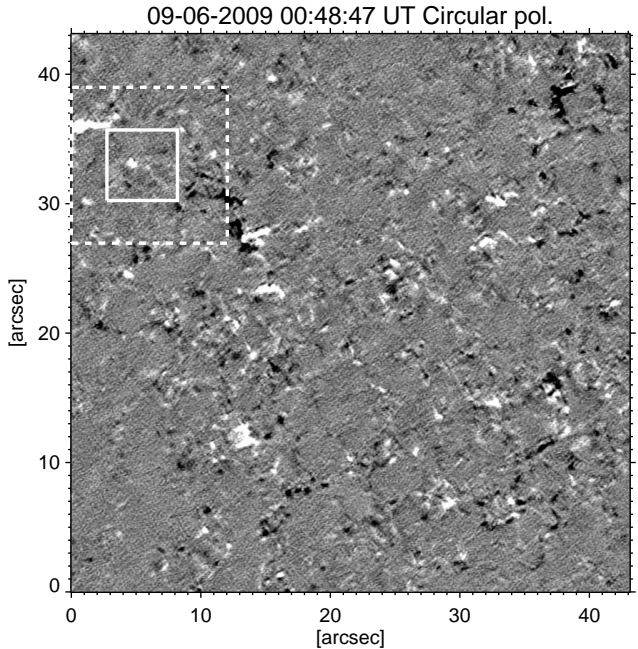}
\caption{Map of the mean circular polarization signal $V_s$ with a scale range of [-1,1]\% of the $I_c$, covering the FOV of IMaX of about 43\arcsec\ $\times$ 43\arcsec. The dashed-line square, with a FOV of 12\farcs 1 $\times$ 12\farcs 1, indicates the location whose continuum-intensity has been aligned by cross-correlating  two consecutive images. The inner solid line square, with a FOV of  5\farcs 5 $\times$ 5\farcs 5, indicates the subregion where the magnetic element is studied in detail, as shown in the sequences displayed in Figures \ref{fig2}, \ref{fig4}, and \ref{fig6}.\label{fig1}}
\end{figure}

We analyze disk center quiet-Sun IMaX spectropolarimetric observations. The data set was obtained on 2009 June 9, 00:36:03--00:58:46 UT. IMaX measured the full Stokes vector in five wavelength positions across the Fe \textsc{i} 5250.217 \AA\ line (Land\'{e} factor g = 3) at $\lambda$ = -80, -40, +40, +80, and +227 m\AA\ from the line center (V5-6 mode). The temporal cadence of a full observing cycle is 33 s, with a pixel size of 0\farcs 055.

IMaX data reduction routines were used for dark-current subtraction, flat-field correction, and polarization crosstalk removal. The blueshift over the field-of-view (FOV) produced by the Fabry-P\'{e}rot interferometer is corrected in the inferred velocity values. The applied restoration technique requires an apodization that effectively reduces the IMaX FOV down to about 43\arcsec\ $\times$ 43\arcsec. The spatial resolution has been estimated to be 0\farcs 15 - 0\farcs 18 (after reconstruction), and the noise level in each Stokes parameter is about 2.5 $\times$ 10$^{-3}$ in units of the continuum intensity. The rms contrast of the quiet-Sun granulation obtained from IMaX continuum data is around 13.5\%, which testifies to the high quality of the IMaX/\textsc{Sunrise} images. For further details about data reduction, we refer to \citet{2011SoPh..268...57M}.

We obtained maps of the mean circular polarization averaged over the line, $V_s$, and of the mean linear polarization signal, $L_s$, given respectively by
\begin{eqnarray}
V_s & = & \frac{1}{4\langle I_{c}\rangle}\sum_{i=1}^4\epsilon_i \cdot V_i, \nonumber \\
L_s & = & \frac{1}{4\langle I_{c}\rangle}\sum_{i=1}^4\sqrt{Q_i^2+U_i^2},  \nonumber 
\end{eqnarray} 
where $\langle I_{c}\rangle$ is the continuum intensity averaged over the IMaX FoV, $\epsilon=$[$1, 1,-1,-1$] and $i$ runs over the first four wavelength positions.  In the weak field regime, $V_s$ very approximately scales with the longitudinal magnetic component, while $L_s$  is a measure of the transverse (horizontal) component of the magnetic field.

We carried out inversions of the Stokes vector observed with IMaX using the SIR code \citep{1992ApJ...398..375R}. This code, based on the Levenberg-Marquardt algorithm, numerically solves the radiative transfer equation (RTE) along the line-of-sight (LOS)  under the assumption of local thermodynamic equilibrium and minimizes the difference between the measured and the computed synthetic Stokes profiles using response functions. 

Using two nodes at which the temperature is explicitely determined, the inversion yields the temperature stratification in the range $-4.0 < \log \tau < 0$ through inter- and extra-polation, where $\tau$ is the continuum optical depth at 5000 \AA. However it is worth noting that with only 5 wavelength points and a single, very temperature dependent spectral line, the temperature is not constrained reliably in layers above $\log\tau= -1.5$ or $-2$ (depending on the type of feature and the strength of the line). We also obtain the height-independent magnetic field strength $B$, the inclination and azimuth angles $\gamma$ and $\phi$, the LOS velocity, and the microturbulent velocity, using one node for each of them. The magnetic filling factor is assumed to be unity and the macroturbulent velocity is set to zero due to the high spatial resolution of the data.

Figure \ref{fig1} displays a map of the mean circular polarization for the FoV covered by the observations, about 43\arcsec\ $\times$ 43\arcsec over a quiet region at disk center. It shows many internetwork flux concentrations along with stronger and larger flux elements probably belonging to the network.

After applying a p-mode subsonic filter \citep{1989ApJ...336..475T} to the continuum intensity and LOS velocity maps, we focus on a smaller area of 12\farcs 1 $\times$ 12\farcs 1, indicated by the white dashed-line square in Figure \ref{fig1}. On this subfield, we aligned the continuum intensity maps by applying a cross-correlation technique on two consecutive images. The same displacement correction was also applied to the other parameters of interest. Finally, we restricted ourselves to study an even smaller area of  5\farcs 5 $\times$ 5\farcs 5, displayed by the white solid-line square in Figure \ref{fig1}. Within this area, we constructed movies of the continuum intensity, LOS velocity, circular polarization, and field strength, and we obtained horizontal velocity maps of the first three parameters time averaged over a given interval by using the local correlation tracking (LCT) technique \citep{1986ApOpt..25..392N} as implemented by \citet{1994IRN..31}. This technique selects small sub-fields around the same pixel in contiguous frames, and correlates them to find the best-match displacement. The sub-fields are defined by a Gaussian tracking window with a full width at half maximum (FWHM) of 0\farcs 3. In order to help the algorithm, the original images are interpolated in time (linearly) and space (bi-linearly) so that the pixel size and cadence is reduced to 0\farcs 028 and 11 s respectively. These interpolations do not add any significant information and therefore do not change the results. They only help to get less noisy velocity maps. An example of such horizontal velocity maps is shown in Figure \ref{fig3} which is discussed later in Section \ref{FluxAdvection}.




\section{Results}

\subsection{Flux Concentration} \label{FluxAdvection}

\begin{figure*}[t]
\includegraphics[angle=90,scale=.75]{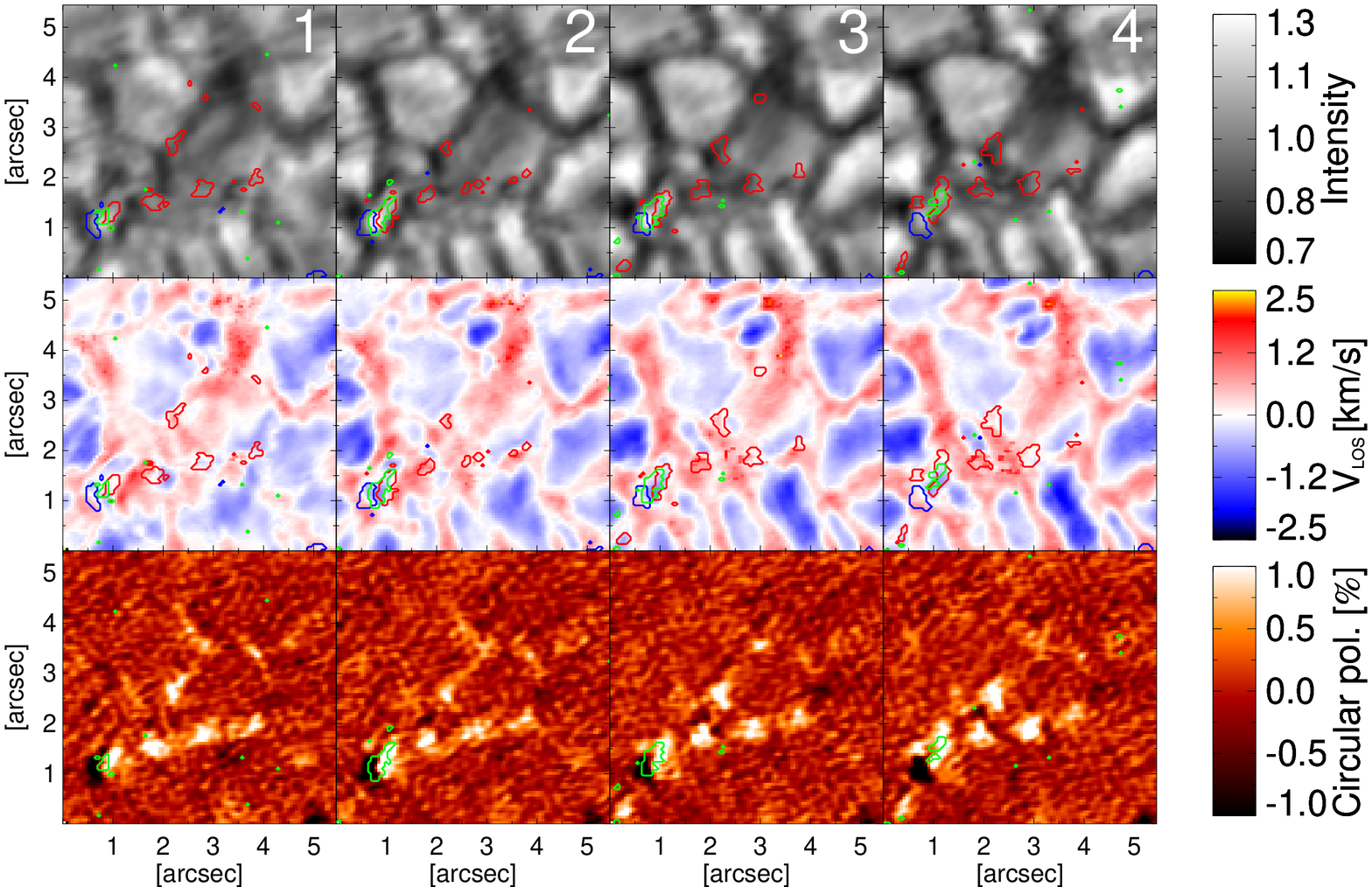}
\includegraphics[angle=90,scale=.75]{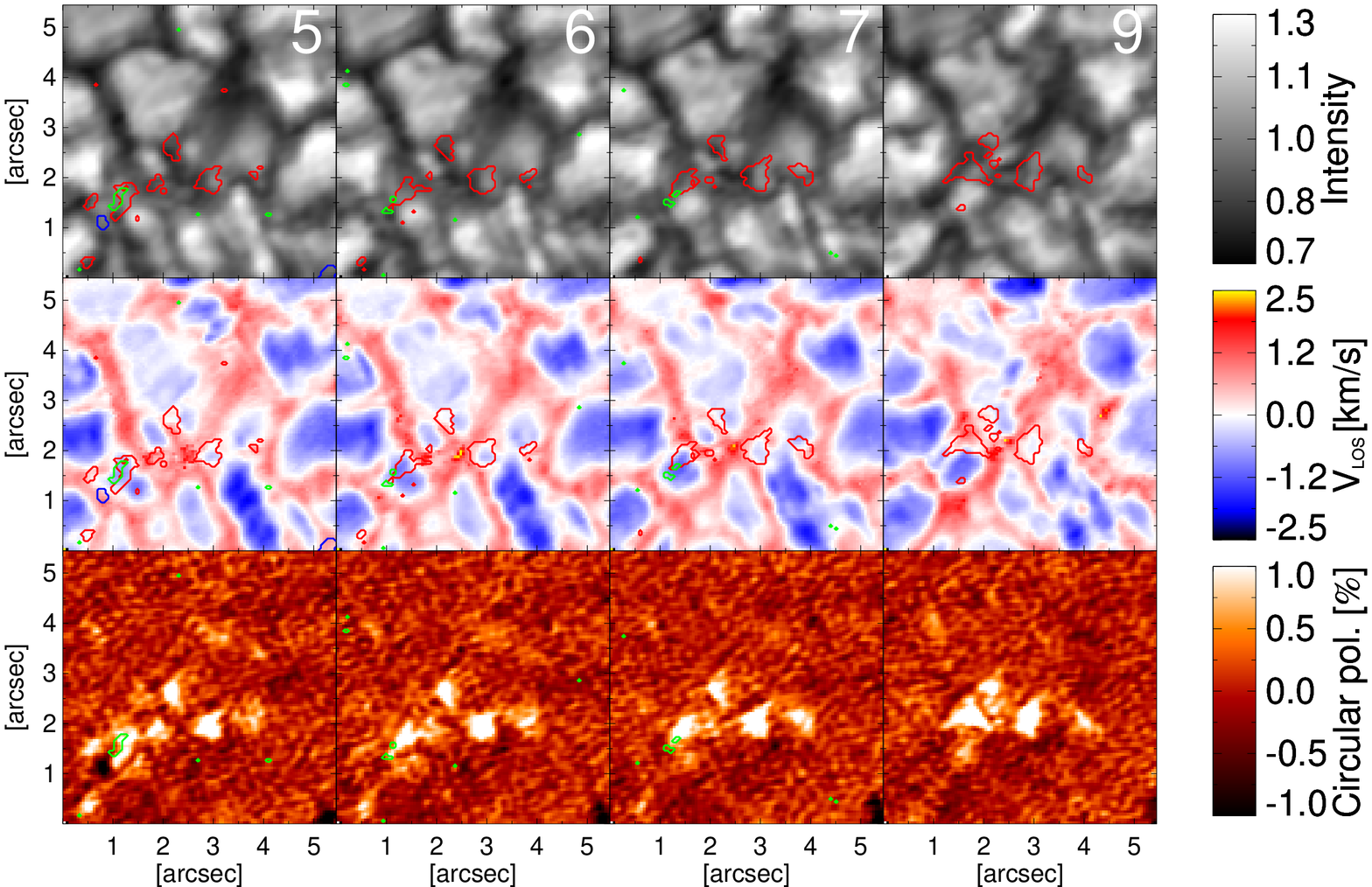}
\caption{Temporal sequence of the continuum intensity maps (top rows), LOS velocity (middle rows), and total circular polarization $V_s$ (bottom rows) during the flux concentration phase. Red (blue) contours over the maps represent a circular polarization signal of $+0.8$ ($-0.8$)\% of the $I_c$, and green contours represent a linear polarization signal of $0.8$\% of the $I_c$. The elapsed time between consecutive frames is 33.25 s (except between the last two frames), and  runs from left to right and continues in the lower set of panels, as numbered in continuum intensity maps. This figure is also available as an animation in the electronic edition of {\it The Astrophysical Journal}.
\label{fig2}}
\end{figure*}

\begin{figure*}[t]
\includegraphics[angle=90,scale=.76]{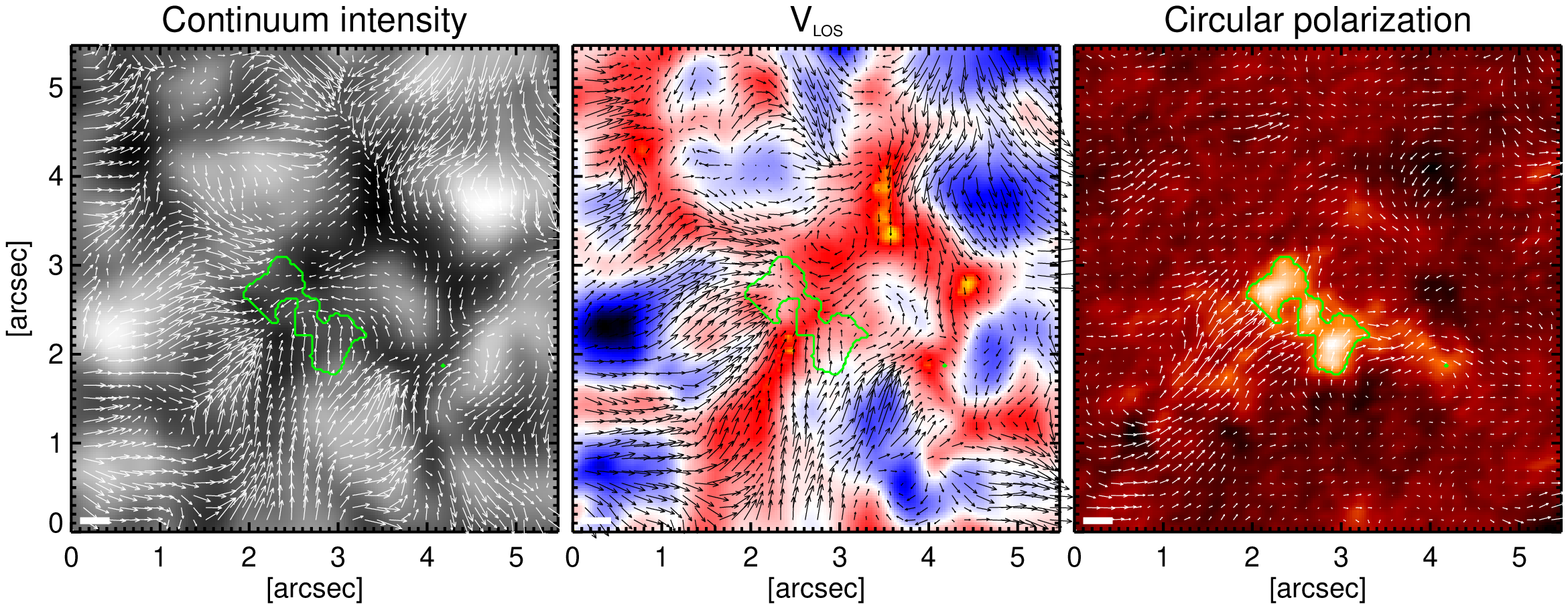}
\caption{Horizontal velocity maps derived through the LCT technique averaged over the flux concentration phase ($\sim 11$ min, frames 1-21). Proper motions of the parameters shown in Figure \ref{fig2} are displayed. From left to right: continuum intensity, LOS velocity, and mean circular polarization. The images are averaged in time over this phase. The length of the white bar at coordinates (0.1,0.1) corresponds to 1.8 km s$^{-1}$. Green contours over the averaged maps represent a circular polarization signal of $+0.5\%$ of the $I_c$.
\label{fig3}}
\end{figure*}

Figure \ref{fig2} displays a temporal sequence of the continuum intensity $I_c$ normalized to the average value over the IMaX FoV, the  LOS velocity retrieved with SIR, and the circular polarization $V_s$ maps. Note that not all frames in the figure are consecutive. The sequence describes the first phase in the magnetic structure's evolution, mostly characterized by the rise of a small-scale magnetic flux loop and the granular dragging of its footpoints to nearby intergranular lanes. The high and constant spatial resolution of the data allows us to trace the dynamics of sub-arcsecond magnetic patches. We use this property to track the advection of polarization signal by the horizontal plasma flows. Red and blue contours encircle areas with positive and negative circular polarization respectively, whereas green contours indicate regions with a significant linear polarization signal. 

In frame 1, at coordinates $[0\farcs 75, 1\farcs 25]$ there is a small-scale loop \citep{2007A&A...469L..39M,2009ApJ...700.1391M,2010ApJ...723L.149D} with a bipole flux of $4 \times 10^{16}$ \textsc{Mx} and a field strength peak of 300 G above a pre-existing granule.\footnote[1]{Unless otherwise stated, fluxes are calculated throughout the paper by considering the area enclosed by contours of $V_{s} = 8\times 10^{-3}$. Field strength values are calculated as averages over 3$\times$3 pixel boxes to reduce the influence of noise.} Two opposite-polarity footpoints are connected by a quite strong $L_{s}$ signal between them. A statistical study of granular scale loops has already been carried out  by \citet{2012ApJ...755..175M}, using these IMaX data. From the same  \textsc{Sunrise} science flight, \citet{2012ApJ...745..160G} reported on the evolution of a larger, intermediate-scale, magnetic bipole. 

Here, the evolution of the loop can easily be followed in the subsequent $V_s$ frames. The footpoints move from within the granule to nearby intergranular lanes. This motion represents a phase of flux expulsion. At the same time, the $\Omega$-shaped loop is rising as witnessed by the progressive disappearance of the $L_{s}$ signal (the loop top) while the footpoints stay in the photosphere. Therefore, the underlying granule not only helps to bring the loop to higher layers but also advects the footpoints to the intergranular lanes. At frame 6, the negative footpoint disappears, which is more likely due to cancellation with an opposite polarity  patch appearing from frame 2 to 5 at coordinates $[0\farcs 75, 1\farcs 5]$, just above the negative footpoint. Note that the weakening of the $L_{s}$ signal occurs when flux is cancelling, hence suggesting that cancellation also contributes to the disappearance of $L_{s}$. This cancellation of opposite-polarity magnetic patches should be related with some form of magnetic reconnection, in a similar way as the strong blueshift events first observed by \citet{2010ApJ...723L.144B} using the same IMaX data. In fact, the supersonic upflow associated with this particular magnetic cancellation event is visible in the beginning of Animations 1 and 3 at coordinates $[5\arcsec, 33\arcsec]$ in \citet{2010ApJ...723L.144B}. These quiet-Sun jets have been confirmed in Hinode/SP data \citep{2011A&A...530A.111M}, and their relation with horizontal field patches have been highlighted by \citet{2013A&A...558A..30Q}.

By the time the positive footpoint moves towards the intergranular lane and the negative one is cancelled, three additional patches of positive $V_s$ have appeared in contiguous convective downflow areas (frame 9 in Figure \ref{fig2}). The whole evolution (frames 1 through 21) of those small magnetic features can be followed in the Animation 1 included in the electronic edition of the journal. As expected from the small sizes of these magnetic patches, they move along intergranular lanes driven by the horizontal displacement of the granules. This dragging of magnetic patches, gives rise to a number of merging and splitting processes, which result in a bigger and stronger magnetic structure at the end of this phase (frame 21 in the animation). As a result, the magnetic element carries a flux of $5\times 10^{17}$ Mx with a field strength peak of 600 G. Note that this and other field strength values given in this paper are lower limits, since the inversion assumes filling factor unity. We can call these later stages of evolution as ``flux concentration by granular advection''.

Figure \ref{fig3} displays the horizontal velocity maps inferred by the LCT and averaged over this phase  ($\sim 11$ min, frames 1-21). The flows derived from the circular polarization show motions of magnetic features. On the other hand, the flows derived from the brightness and the LOS velocities show the evolution of granulation with time. The Gaussian tracking window with a FWHM of 0\farcs 3 could allow us to infer flows at a sub-granular scale. However, as we are averaging over a time period (21 frames, 11 min) much longer than that expected for the lifetime of internal convective velocities, the evolution of granules is the dominant contributor to these horizontal flows.  This is supported by  \citet{2013A&A...555A.136V}, which presented rigorous testing of  LCT algorithm by comparing its results with velocities in an MHD simulation. In particular, they found that proper motions of single granules are well captured when flow maps are averaged over 15 and 30 min, and claim that even with very narrow sampling windows and short time cadences, recovering details of the plasma flows might be unreliable.

The proper motions of continuum intensity and LOS velocity illustrate how granules converge towards the center of the maps where the strongest magnetic flux concentration is found. The circular polarization map shows that the small magnetic patches also move in the same direction as the granules, so that the magnetic flux is concentrated in the center of the map. Any motion of magnetic features then simply implies a motion caused by the evolution of granulation. Thus, the magnetic features can stay within the lanes all the time and still move dragged by these flows. Therefore, Figure \ref{fig3} is indicative of flux concentration by granular advection.

In summary, we can say that this phase encompasses four clear stages, namely, a) the rise of an $\Omega$ loop within a granule; b) the expulsion of footpoints towards nearby intergranular lanes; c) flux cancellation of one of the footpoints with an opposite-polarity patch, likely through a reconnection process; and d) the increase of flux in the other footpoint by merging with pre-existing patches of the same polarity, driven by granular advection.

\subsection{Convective Collapse} \label{ConvectiveCollapse}

\begin{figure*}[t]
\includegraphics[angle=90,scale=.75]{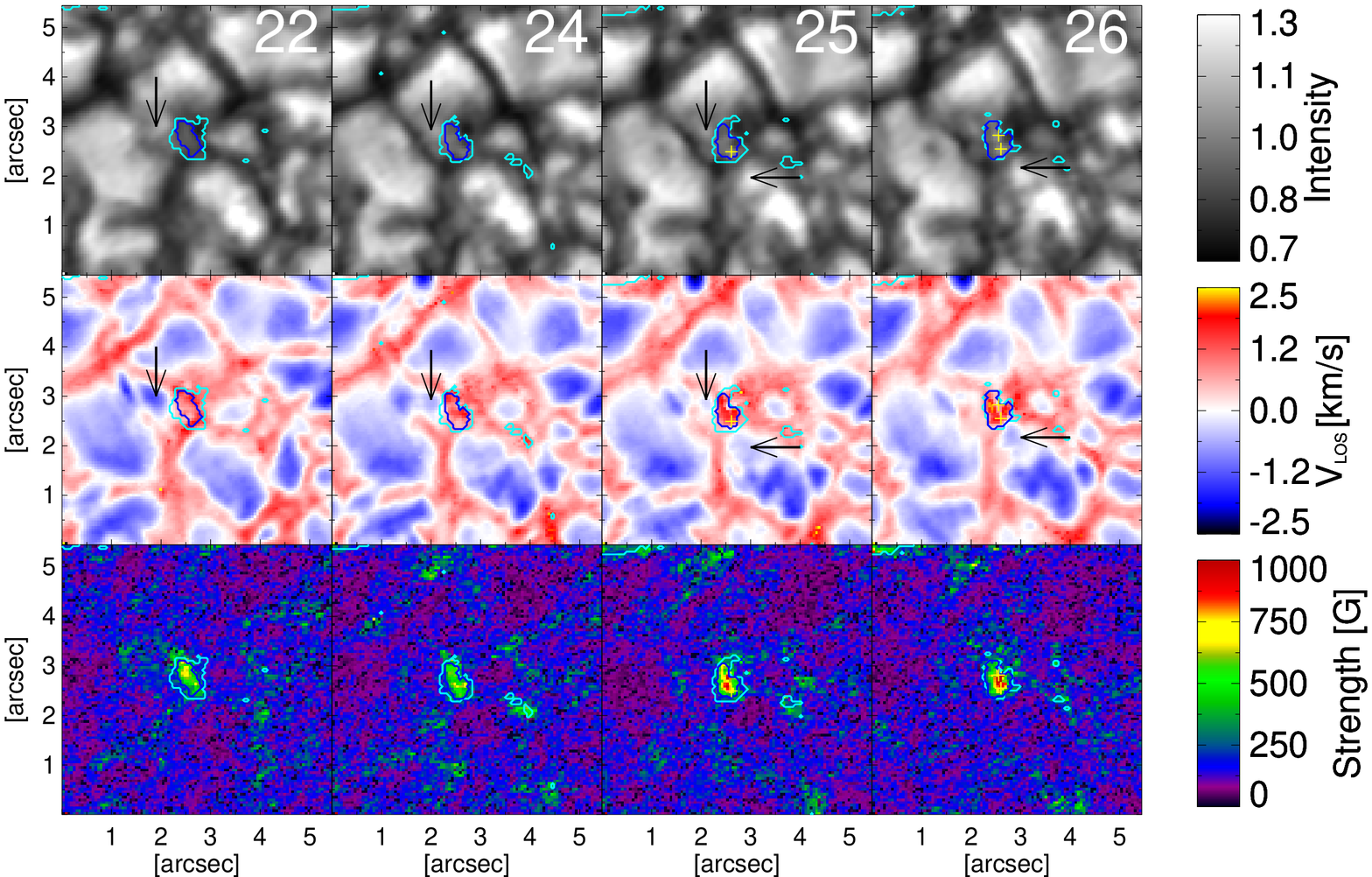}
\includegraphics[angle=90,scale=.75]{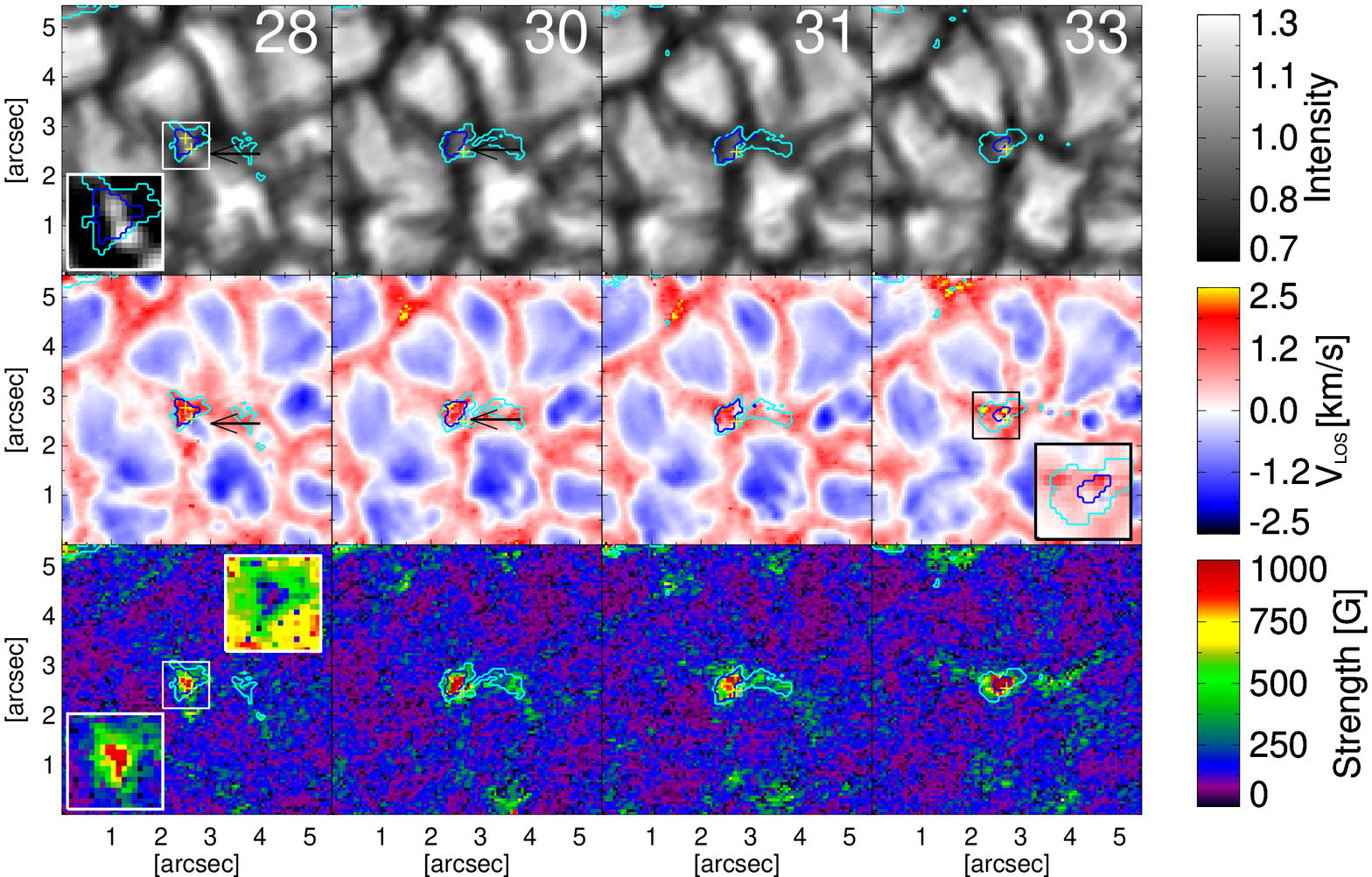}
\caption{Same as Figure \ref{fig2}, during the convective collapse phase. Note that here the bottom rows show field strength maps instead of circular polarization. Cyan line represents iso-magnetic flux density contours of $1 \times 10^{15}$ Mx. Blue line delineates regions containing a time-constant magnetic flux of $3.5 \times 10^{17}$ Mx. The yellow cross marks show the bright points location, and the arrows point to small-scale converging upflow features. White boxes in frame 28 display a zoom of the magnetic element for continuum intensity (saturated to [0.9,1.1] $I_{c}$), field strength and inclination; the latter is saturated to [20,120] degrees. The black box in frame 33 displays a zoom of the LOS velocity saturated to [-5,5] km s$^{-1}$. This figure is also available as an animation in the electronic edition of {\it The Astrophysical Journal}.
\label{fig4}}
\end{figure*}

\begin{figure}[t]
\includegraphics[scale=1]{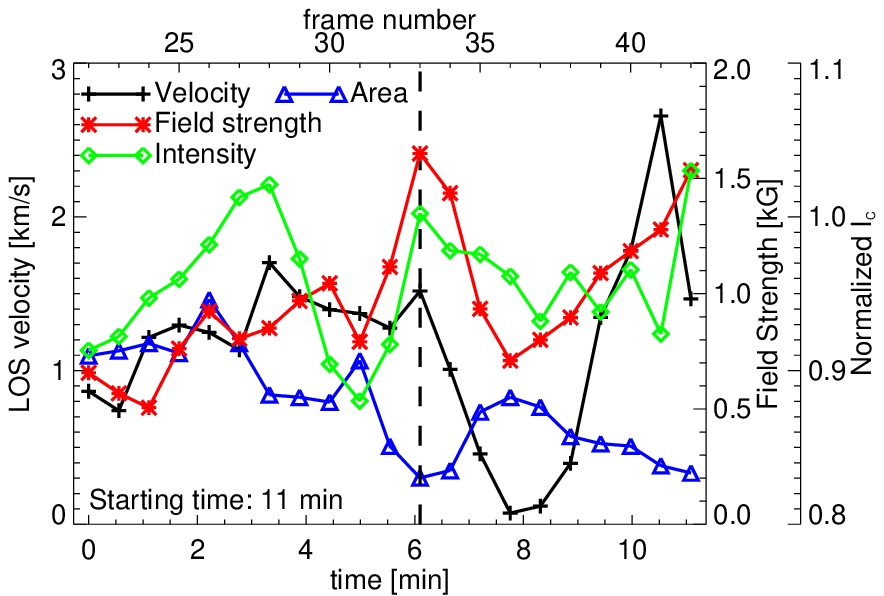}
\includegraphics[scale=1]{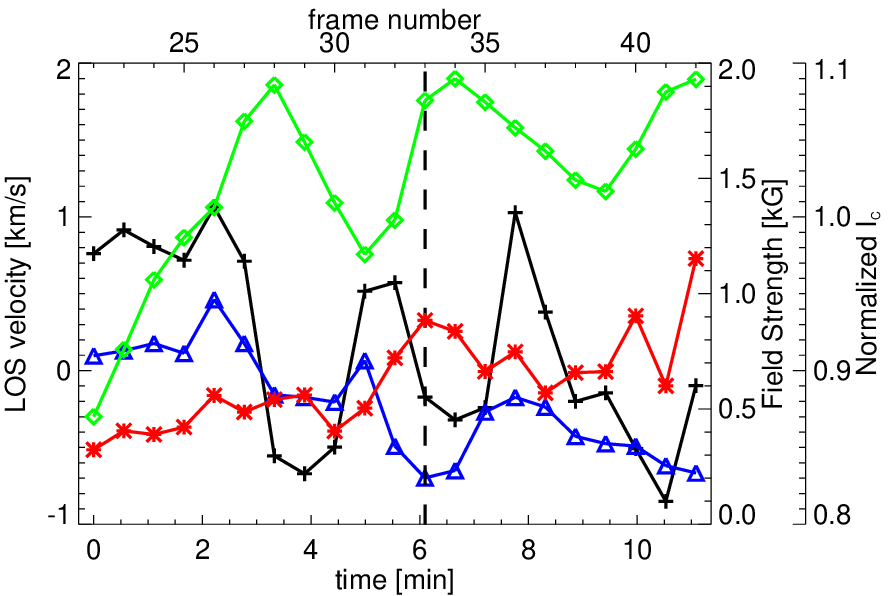}
\caption{Evolution of LOS velocity (black line with plus symbols), field strength (red line with asterisks), and continuum intensity (green line with diamonds), of the flux tube core (Figure 5a) and one of the bright points (Figure 5b), for frames 22-42.The flux tube core is defined by the blue contour shown in Figures \ref{fig4} and \ref{fig6}, which contains a constant magnetic flux of $3.5 \times 10^{17}$ Mx. We display the evolution of the flux tube core area (blue line with triangles) in both a and b panels. The area is measured in Mm$^{2}$, and we use the same y-axis as the one for LOS velocity. Note that the area has been multiplied by a factor 10 for better visualization. In addition, a value of 1 is subtracted from it in panel b. The physical parameters are extracted by averaging over 9 pixels around the centroid of the magnetic core and bright point for the panels respectively. The dashed vertical line indicates the end of the convective collapse and the beginning of an oscillation phase. Time 0 in the x-axis corresponds to 11 min after the observations started, as marked in the Figure. The upper x-axis at the top of each panels mark the frame numbers as shown in Figures \ref{fig4} and \ref{fig6}.
\label{fig5}}
\end{figure}

Figure \ref{fig4} shows the formation of a small kG flux concentration. In frame 22 there is a small magnetic patch formed during the flux concentration phase described above. The magnetic structure has field strengths of about 400 G with a maximum value up to roughly 600 G in its center. These strengths are of the order of the typical equipartition field strength (300-500 G) for granules. Note that only eight of the total twelve frames of this convective collapse phase are shown. From top to bottom, rows correspond to continuum intensity, LOS velocity, and magnetic field strength. Overplotted are contours defining the magnetic element. The innermost, blue one corresponds to a region of constant magnetic flux. The external, cyan contour marks regions with longitudinal field components stronger than 60 G. These two contours will be used until the end of the present study. In the three bottom rows, inserts display zooms of the little squares containing the magnetic structure near the centres of the frames. In the case of the velocity insert in frame 33, the scale is doubled (from -5 to +5 km\,s$^{-1}$) in order not to have saturated colors. Two inserts are plotted for the magnetic field strength in frame 28: the bottom one shows a blow-up of the field strength while the top one displays a map of the magnetic inclination in order to illustrate that indeed the magnetic element resembles a fully resolved, canonical flux tube where and almost vertical ($\sim 20\deg$) inner core is surrounded by more inclined ($\sim 70\deg$), canopy-like magnetic fields. The outer, 60-G contour is mostly used to illustrate how very small magnetic patches, external to our main structure at the beginning of this phase, progressively increase in size and are advected by granules until they merge with our structure in frame 31.

Let us concentrate now on the constant-flux region enclosed by the inner contour. It encloses a magnetic flux of $3.5\times 10^{17}$ Mx during this evolution phase. As shown in the first frame, this magnetic patch is embedded in an intergranular lane. As time goes on, the area enclosed by this contour decreases sharply while downward motions and field strengths increase within it until kG fields are reached. To quantitatively analyze the evolution in detail, we select the magnetic core of the structure as the centroid of field strengths within the above-mentioned constant-flux contour. The upper panel of Figure \ref{fig5} displays the LOS velocity (crosses and black line), the magnetic field strength (asterisks and red line), and the continuum intensity (diamonds and green line) of such a magnetic flux tube core and the area (triangles and blue line) of the constant flux region. To increase the S/N in the magnetic core physical parameters, we represent averages over the core itself and its eight surrounding pixels. Although not explicit in the axis legends, areas are measured in Mm$^{2}$ and multiplied by 10 so that they can be read with the same scale as LOS velocities. Labels in the upper horizontal axis corresponds to frame numbers. The vertical dashed line corresponds to the end of this phase, at frame 33, 6.1 min after its start. At this moment the magnetic field reaches strength up to  1.6 kG, compared with the initial 600 G, while the downflow has grown from 0.9 km {s}$^{-1}$ to 1.5 km s$^{-1}$. To estimate the noise-induced uncertainty in the field strength and LOS velocity, we repeated the inversions with 100 different realizations of added noise to the observed Stokes profiles. Amplitudes of $10^{-3}\, I_{c}$ were used. The standard deviation of the 100 results is 100 G and 100 m s$^{-1}$. Note that the area of the whole magnetic structure runs in almost anti-phase to the strength of the magnetic core, decreasing from 0.11 to 0.03 Mm$^{2}$. This indicates that magnetic flux is conserved and that the flux contribution of the canopy fields is not very significant, as expected.

In general, the continuum intensity also seems to gradually increase as the field strength intensifies. However, in this case, there is no clear correlation, i.e, the peak intensity is reached before the field strength has attained its maximum. The change in brightness will be studied in more detail in Section \ref{BrightFeature}.

\subsubsection{Converging granules and small-scale upflow features} \label{UpflowFeatures}

From the LOS velocity maps in Figure \ref{fig4}, the presence of two small-scale upflows at the periphery  of our magnetic element are evident. These upflows lie above the estimated uncertainties. We mark them using small arrows in the $I_c$ and LOS velocity maps. 

The first of those features is indicated by a downward arrow from frame 22 to 25. It is a small upflow fragment detached from a  bigger granule. It cools down quickly as it converges towards the magnetic structure. It completely disappears before the strongest downward velocities can be observed in the interior of the flux concentration from frame 26 on.

The feature indicated by the left pointing arrow, in frames 25-30, begins  as a typical granule-shaped upflow located close to the magnetic structure. As it evolves, it splits in two. The fragment closer to the magnetic element starts to converge towards the flux concentration, while the other half shrinks and brightens. The converging fragment breaks up as it ``collides'' against the magnetic flux tube. These fragments are also seen in continuum intensity maps. In neither case do we observe the penetration of the features into the magnetic element as they do in \citet{1999ApJ...511..436S} for pores. In addition to these fragmenting processes, the granules as a whole continue converging and dragging small magnetic patches towards the center of the map where the magnetic flux concentration is located. In frames 30 through 33, one can see that small upflowing fragments detached from the upper central and upper right granules to our structure approach the small magnetic feature while it coalesces into our magnetic element as we commented on in Section \ref{ConvectiveCollapse}. 

Besides the above mentioned convergence, the shapes of the granules get perturbed by the presence of magnetic fields. In particular, they lengthen in the direction of the magnetic element and the new weak magnetic feature, thus adopting a ``petal-like'' appearance whose sharpest corner points towards the magnetic tube. All together, and surrounding the magnetic element, a characteristic ``daisy-like'' granular pattern \citep{1989SoPh..119..229M, 1992SoPh..141...27M} is observed at the end of the phase (frame 33 in Figure \ref{fig4}).

The sum of the above observations makes it seem as if the flux tube behaved as a sink, which attracted the surrounding convective upflows. This effect is clearly observed in Animation 1, included in the electronic edition of the journal. There are two effects that may contribute to this seeming attraction. Firstly, magnetic elements provide a larger surface area through which radiation can escape and hence the surrounding gas  be cooled \citep{1976SoPh...50..269S}. This leads to more vigorous convection \citep{1984A&A...139..435D}. Secondly, this magnetic element is located at the intersection of a number of granules, where convective downflows are often particularly strong, so that horizontal flows tend to go towards them. Also, magnetic elements are often located near the centres of vortices that pull the nearby granules towards them \citep[e.g.][]{2008ApJ...687L.131B,2010ApJ...723L.139B}, which may also contribute.

\subsubsection{Bright points inside the flux tube} \label{BrightFeature}

\begin{figure*}[t]
\includegraphics[angle=90,scale=.75]{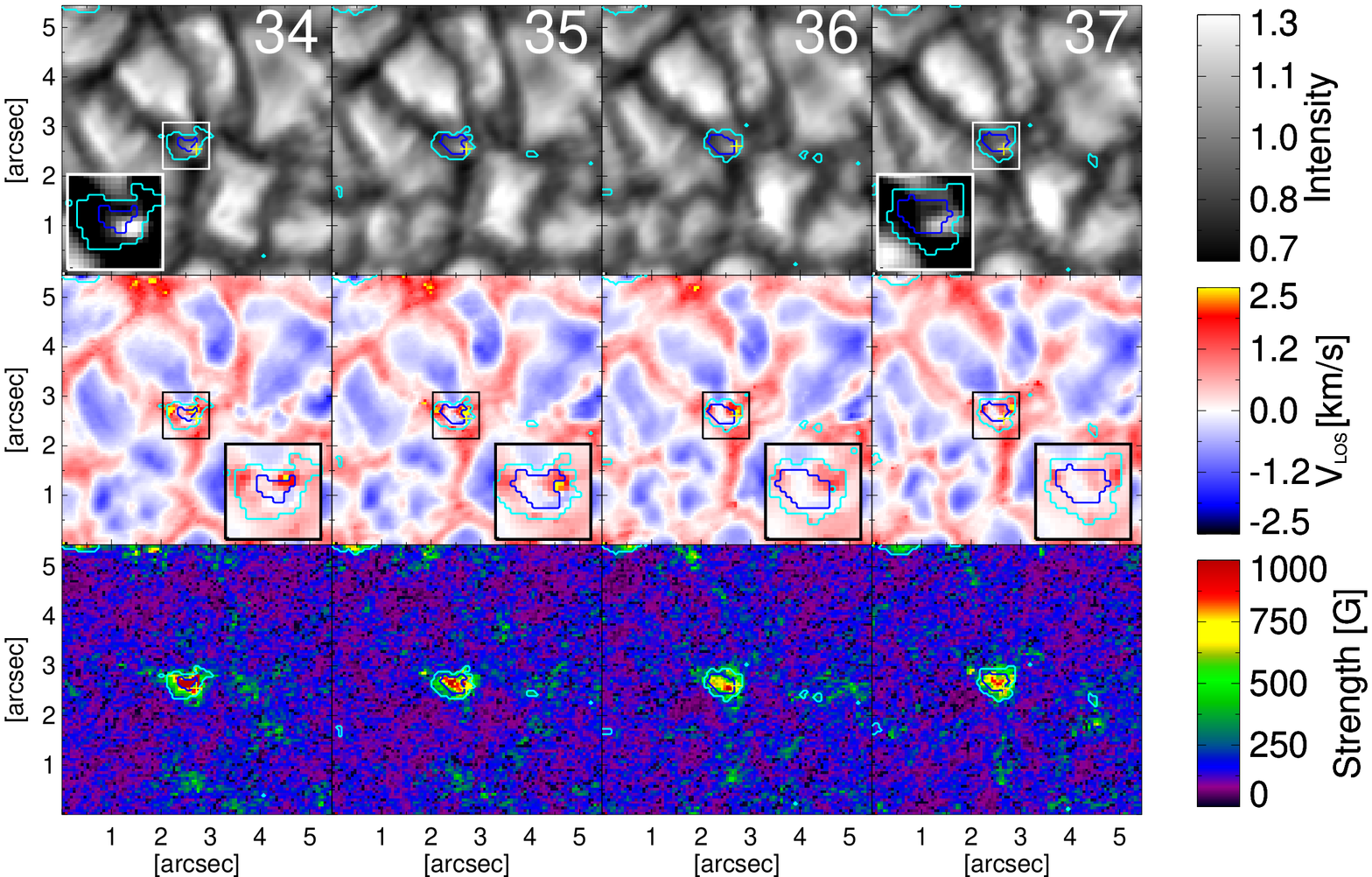}
\includegraphics[angle=90,scale=.75]{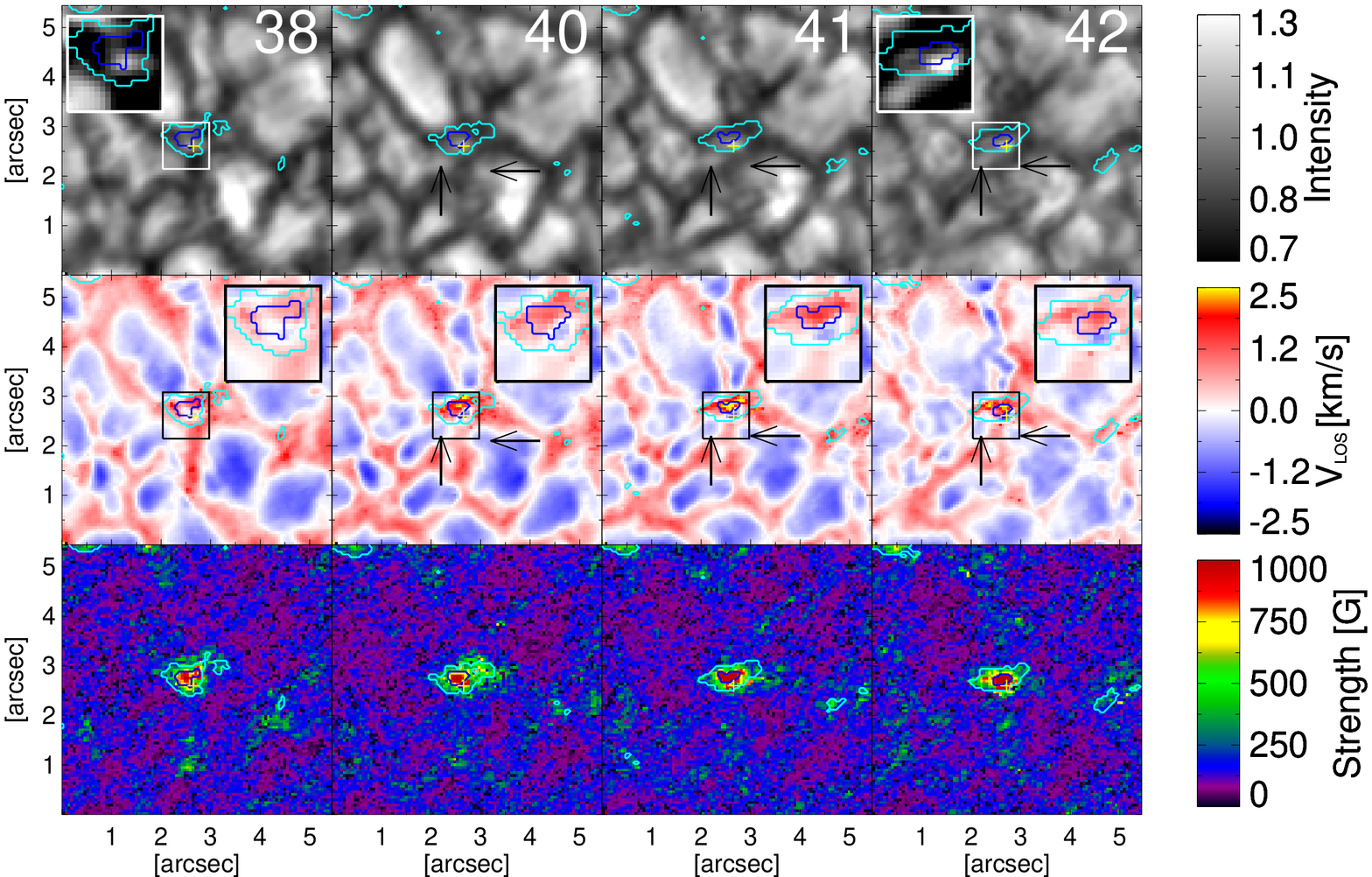}
\caption{Same as Figure \ref{fig4}, during the evolution of the mature flux tube. This figure is also available as an animation in the electronic edition of {\it The Astrophysical Journal}.
\label{fig6}}
\end{figure*}

On the continuum intensity maps in Figure \ref{fig4}, two small-scale BPs are glimpsed inside the magnetic flux tube. The BPs appear as the downflow and field strength increase in the interior of the magnetic element and they are indicated by two yellow plus symbols. They can best be discerned in the insert zoom of the continuum intensity map of frame 28, the scale ranges from 0.9 to 1.1 $I_{c}$ in order to enhance their contrast. It is worth noting that these BPs are smaller than the magnetic element. Hence, unlike the usual assumption, neither one can be identified with a single flux tube.

The BPs are first located close to the core of the magnetic element (frame 26) and, consequently, can be associated with the flux tube's evacuation. Although not exactly coincident with the core of the magnetic element, the uppermost one is close enough for its brightness evolution to be responsible for most of the continuum intensity variation we see in Figure \ref{fig5}a. Until the BP disappears in frame 29, its brightness correlates fairly well with the magnetic field strength intensification. Note, however, that at this point in time the field is still comparatively weak, since this is still prior to the convective collapse. The disappearance is almost simultaneous to the merging of the main magnetic element considered here with smaller magnetic patches mentioned in the first paragraph of Section \ref{ConvectiveCollapse}. The second BP (the one closer to the canopy) remains observable until the end of the time series. Thus, we can follow the evolution of its continuum intensity, LOS velocity, and magnetic field strength as calculated from an average of the nine-pixel box centered on the BP brightness centroid. Such an evolution is displayed in Figure \ref{fig5}b. Starting from frame 25, the BP gets closer and closer to the lower edge of the magnetic element as the granules and small granular fragments converge on the magnetic structure. The lowermost upflow feature described in Section \ref{UpflowFeatures} arrives at the same time to that edge, thus producing a reversal in LOS velocity. Simultaneously, the continuum intensity reaches its peak (1.1 $I_c$; frame 28). Later, the intensity starts to decrease as the small upflow breaks up and the magnetic element merges with the neighboring weak magnetic feature (frames 31-33; see  Figure \ref{fig4}). Following this,  a downflow (0.5 km s$^{-1}$) is re-established, but not for long because a narrow, weak upflow plume appears at the location of the BP by the end of the intensification phase at frame 33 (see insert). We cannot say whether this upflow feature is actually a part of the magnetic element boundary or just a non-magnetic gas parcel below the tube canopy, but with its emergence, the continuum intensity increases sharply, reaching again values about 1.1 $I_c$. We speculate that the presence of hot (bright) gas next to the magnetic element leads to an intensification of the bright point near the edge of the magnetic feature, since this hot gas heats up and brightens the wall of the magnetic element.

\subsection{Mature Flux Tube} \label{MFT}

Figure \ref{fig6} shows the newly formed kG magnetic flux concentration. The evolution of the different parameters at the flux tube's core can be followed in Figure \ref{fig5}a after the vertical dashed line.

Interestingly, rather than keeping a constant magnetic field strength, it drops below 1 kG where it stays for a number of frames before shooting up again. Oscillations of the strength of quiet-Sun magnetic fields have been observed for the first time by \citet{2011ApJ...730L..37M} using the same IMaX data. They detected those oscillations by studying the changes with time of the area enclosed in a contour containing a constant magnetic flux. Here we have defined the area in the same way and, as expected, it varies in anti-phase with the field strength, whereas the LOS velocity is in phase with the field strength. First, it decreases to 0 km$\,$s$^{-1}$ while the field strength drops to 700 G, and then it grows to 2.8 km$\,$s$^{-1}$ as the field strength intensifies to 1500 G. Meanwhile, the continuum intensity remains almost constant in the interior of the magnetic core, around 1.0 $I_c$. However, the intensity of the second BP shows a related oscillatory-like behavior at the wall of the magnetic element (see Figure \ref{fig5}b after the vertical dashed line), reaching large values at times when the field strength is also large.
 
We have already mentioned the emergence of a small upflow plume at the end of the convective collapse (when the area is smallest). As shown in Figure \ref{fig6}, the following evolution of the BP is closely associated with that of the small upflow. Note that the upflow is observed exactly at the location where the BP is present. This fact can also be seen quantitatively in Figure \ref{fig5}b, after the vertical dashed line. As the upflow weakens, the gas cools down, the area increases and the intensity is reduced (time steps 34-36). Soon after that (time step 38) the area of the magnetic element starts to decrease again and a second upflow plume is detected followed by a rise in intensity. 

The second upflow feature appears when the magnetic element is being compressed (area reduced) again by the converging surrounding granules. Similar to those described in Section \ref{UpflowFeatures}, two small-scale upflow features are again detected moving towards the flux concentration (indicated by arrows in Figure \ref{fig6}, frames 40-42). From the corresponding $I_c$ maps, it can be concluded that they are associated with the splitting of neighboring granules. In addition, the daisy-like appearance is again enhanced.

Furthermore, the emergence of the small upflow takes place while a strong downflow of up to 2.8 km s$^{-1}$ develops in the interior of the flux tube. Correspondingly, an upward/downward velocity pattern is observed within the magnetic element (frames 40-42). Similar small-scale upflow features often surrounded by ring-shaped downflows have already been observed in active plage regions \citep{2010A&A...524A...3N}. They detect them in large structures rather than in isolated BPs, which mostly show downflows, and interpret them as part of a small-scale magneto-convection in the interior of a strong plage solar magnetic field.

The anti-phase behavior of the velocity and brightness in the brightest point of the magnetic feature, suggests that at least some of the continuum brightness enhancements in the magnetic element are related to the presence of flows within it. We did not find any previous mention of such a relation between brightness and LOS velocity in BPs. This can be understood in terms of magneto-convection, with upflows bringing hot gas from below to cool and radiate away at the solar surface. But equivalently, it could also be explained by (magneto-acoustic) waves. Indeed, upwardly propagating acoustic waves are ubiquitous in quiet-Sun magnetic bright points and 3-D MuRAM simulations \citep{2012ApJ...746..183J}. Whatever the mechanism may be, the intensity at the core of the magnetic element stays almost constant, with values close to 1.0 $I_c$, while the BP brightness oscillates with an amplitude of 0.1 $I_c$ as the small upflow features evolve.

\subsubsection{Downflow plumes} \label{DowflowJets}

As soon as the mature kG magnetic element is formed, and in agreement with \citet{1999ASPC..184...38S}, two strong downflow plumes start to be clearly visible at the edge of the magnetic element in Figure \ref{fig6} (frames 34 and 35). As an example, the rightmost  downflow has a mean value of 2 km s$^{-1}$ with speeds roughly up to 6 km s$^{-1}$ (hence, almost supersonic) during its evolution. (Remember that the zoomed areas of the LOS velocity map are scaled to [-5,5] km s$^{-1}$.) These very high speed values have to be taken with caution because of the poor sampling of the spectral line in our data. Nevertheless, the Stokes profiles at these points display significant Doppler shifts although their quantitative value may be more uncertain than those for slower downflows. The downflows get weaker after two minutes, but they strengthen again around the same location at the end of the time sequence. Such strong downflows have been predicted by \citet{1998ApJ...495..468S} in 2-D models of magnetic flux sheets. In their 2-D simulations, the downflows are fed by horizontal flows, and they evolve into ``jets'' as they become narrower and accelerate with depth. Here, the downflows appear in front of elongated converging granules. Accordingly, the gas required for producing such strong and narrow downdrafts is likely provided by these granules.

\subsection{Averaged History} \label{AH}

\begin{figure*}[!ht]
\includegraphics[angle=90,scale=.76]{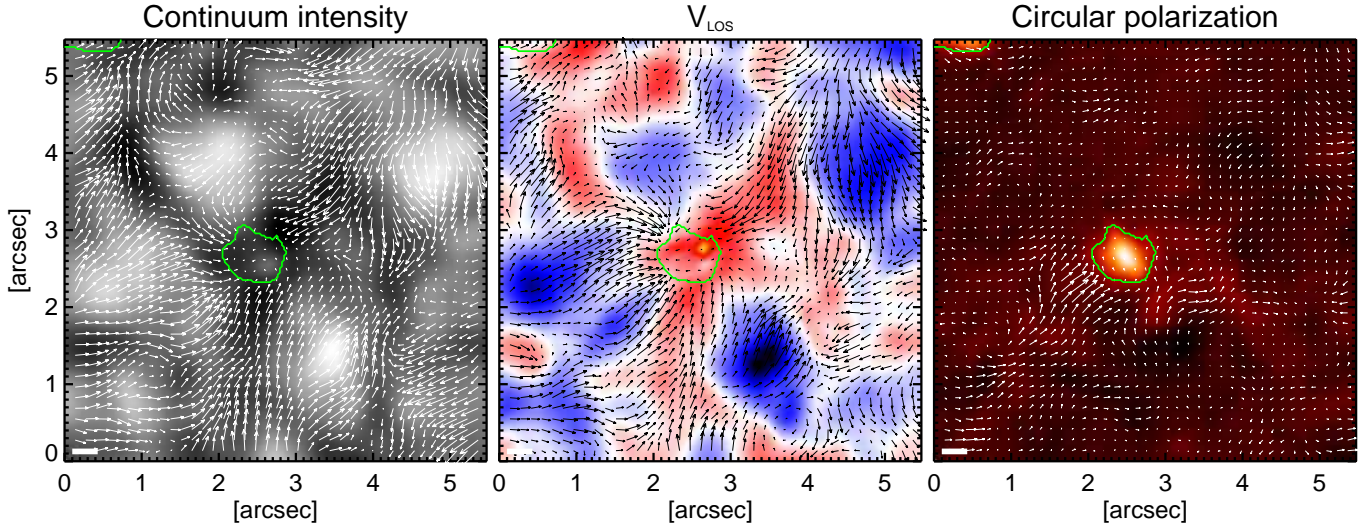}
\caption{Same as Figure \ref{fig3}, averaged over the whole data set ($\sim 23$ min, frames 1-42). This figure is also illustrated by an Animation 1 in the electronic edition of {\it The Astrophysical Journal}. 
\label{fig7}}
\end{figure*}

In order to provide a global picture of the flux tube's history, in Figure \ref{fig7} we display the horizontal velocity maps averaged over the whole time series  ($\sim 23$ min, frames 1-42). Figure \ref{fig7} is also illustrated by an animation where we show movies of continuum intensity, LOS velocity, circular polarization, and field strength maps. Making use of the LCT horizontal velocities we track the advection of passive tracers (corks) initially spread out all over the FoV \citep{1988ApJ...327..964S}.

The horizontal velocity obtained from the continuum intensity and LOS velocity point toward the magnetic feature near the centre of the FOV in Figure \ref{fig7}. A sink is centered at or close to the magnetic element throughout the data set, as persistent flows pointing towards it can be seen after averaging for 20 min \citep[several times the life time of a granule; see][]{2013A&A...555A.136V}. Note that the flows tracked by LCT mainly show the evolution of the granulation with time, so that the converging flows imply that granules and granular fragments converge toward the center of the map. On the other hand, the circular polarization shows the advection of the sub-arcsecond magnetic patches, as described by the flux concentration phase. 

The cork movies shown in the Animation 1 are very interesting. As time goes by, the corks flow towards the structure. While the continuum intensity tracers penetrate the magnetic feature, the ones for the LOS velocity end up at its border. The continuum intensity corks concentrate at two different inner borders of the structure. The location of the BPs described in Section \ref{BrightFeature} match very well with those two concentrations. Furthermore, the LOS velocity corks show two accumulation points at two opposite edges of the structure. The downflow plumes described in Section \ref{DowflowJets} are observed close to if not within these accumulation points.  
 
\section{Summary and Conclusions}

We have presented high resolution observations of the formation and evolution of an isolated quiet-Sun magnetic element and its interaction with the neighbouring convection. We have analyzed the polarization maps and used the SIR inversion code to retrieve LOS velocities and the vector magnetic field.

The history of our magnetic element starts with a small-scale magnetic $\Omega$ loop emerging in a granular upflow. The footpoints are dragged out into nearby intergranular lanes where some pre-existing, sub-arcsecond, positive circular polarity patches are already present. The linear polarity feature disappears at the same time as the negative footpoint cancels with one of those positive polarity patches. This cancellation is associated with a supersonic magnetic upflow detected by \citet{2010ApJ...723L.144B}, which is probably a signature of magnetic reconnection between the cancelling opposite polarity magnetic features. The positive polarity footpoint and the pre-existing flux patches are swept along the lanes by converging granules and concentrated roughly up to (and possibly even beyond) the equipartition field strength (300-500 G).

This process is unable to concentrate the magnetic field significantly above equipartition values. Further intensification is achieved when downdrafts inside the magnetic field concentration are enhanced. According to the canonical convective collapse picture, at this point the tube is evacuated and the flux is compressed by the excess pressure of the surrounding gas. This compression leads to a reduction in the area of the flux concentration and an enhancement of its field strength. This phase of convective collapse is qualitatively consistent with the results of 3-D MuRAM \citep{2005A&A...429..335V} simulations and their comparison with Hinode/SP observations by \citet{2010A&A...509A..76D}. During this process, and due to the formation of a Wilson depression \citep{1976SoPh...50..269S}, the nearby gas cools and hence has a reduced pressure, creating a horizontal pressure gradient with respect to the gas that is located further away. Driven by the pressure gradient that gas then flows towards the magnetic feature. The traces of this inflow can be seen in Figure \ref{fig7}. The field lines act as a stiff hindrance to granular convection, so that  a characteristic daisy-like granular pattern forms, in agreement with the observations of \citet{1989SoPh..119..229M} and \citet{1992SoPh..141...27M}. 

As the flux tube gets cooler than its surrounding at a given geometrical height, it is irradiated laterally from the gas in its immediate surroundings, which is fed by the converging granules and granular fragments. We observe the formation of two BPs within a seemingly single magnetic element.\footnote[2]{Notice that in frame 25 of Figure \ref{fig4} the magnetic element displays two close cores with strengths above 1 kG that subsequently merge in frame 26. This is a phenomenon that can be better seen in other magnetic structures of the same data set and whose study and discussion is deferred to a separate paper.} Both are located at or close to the boundary of the kG feature. We trace the bright point located at the lower edge of the magnetic element using the same approach as \citet{2008ApJ...677L.145N} and find that at the end of the convective collapse phase it displays an upflow. A similar upflow has also been seen by \citet{2001ApJ...560.1010B}. They interpret it as a ``rebound'' arising when the internal downflows turn into upflows, and associate them with the destruction of the flux tube. However, in our study, the small-scale upflow feature does not destroy the magnetic flux tube and rather a large-amplitude variation in area and field strength is observed which may be part of an oscillatory pattern (only a single period is seen, due to the limited length of the observation).

We find that the field strength varies in anti-phase with the area enclosed by a contour of constant magnetic flux, supporting the conclusion drawn by  \citet{2011ApJ...730L..37M} that oscillations in this area are proportional to oscillations in field strength. In the tube core LOS velocity also changes in anti-phase with area. In the BP case however, brightness also varies in anti-phase with area while LOS velocity does it in phase. The BP follows the evolution of an emerging upflow plume. The upflow dissolves as the area increases in size, and a second upflow appears while the area recovers its initial value. Through this evolution the BP oscillates with an amplitude peak of 0.1 $I_c$ above the almost constant brightness intensity at the tube core of about 1 $I_c$. 

Unfortunately, the data-set limited time span does not allow us to know whether an oscillations will continue or not. However, the magnetic field oscillations detected by \citet{2011ApJ...730L..37M}, as well as observations of BPs experiencing several brightness enhancements during their life \citep{1992SoPh..141...27M}, suggest that the magnetic element could undergo more oscillations.

Once the mature flux tube has been formed, we also find narrow, strong downflows at it edges. Two-dimensional models by \citet{1998ApJ...495..468S} predict flux sheets bordered by narrow downflows. The classical picture for the creation of asymmetries in the presence of canopies \citep{1988A&A...206L..37G,1993SSRv...63....1S} predicts the appearance of such downflows. \citet{1997ApJ...478L..45B, 2000ApJ...535..489B} already detected them in unresolved magnetic flux tubes. Here, we do not find rings of downflows bordering  the magnetic structure, but rather downflow plumes at the edge's of the flux concentration, similar to those observed by \citet{2004ApJ...604..906R} in active region flux tubes. These downflows are accompanied by small-scale upflow features that appear at the external border of the magnetic element core. Our high spatial resolution findings agree very well with those obtained by \citet{2012ApJ...758L..40M}, also based on IMaX data.

Our new observation of a strong anti-phase correlation between the continuum intensity and the LOS velocity within the BP deserves a small discussion. A first suggestion can be drawn out of it: at least part of the continuum brightness of the magnetic feature is related to the presence of flows within it. Most notably, the brightness enhancement and the anti-phase velocity are not seen in the central, strongest core of the magnetic element. Rather they appear at well-localized, small-scale places of its external part. Whether this phenomenon is common to this type of elements or not cannot be ascertained from our present data. Further studies focused on this topic promise to offer new insight into the physics of quiet-Sun magnetic flux tubes. 

Several mechanisms can be invoked to explain the possible oscillatory behavior. Overturning magneto-convection could be one such mechanism as the detailed studies by, e.g., \citet{1996MNRAS.283.1153W} suggest. Another possible scenario is provided by overstable oscillations \citep{1979SoPh...61..363S,1985A&A...143...39H} that can start as soon as the collapse has stopped. If the pattern we observe is not of a convective origin, we then have a third possibility: we might be witnessing the upward propagation of acoustic waves \citep{2012ApJ...746..183J}. If so, waves could be excited through magnetic pumping \citep{2011ApJ...730L..24K}. Indeed, transient downflows in the immediate surroundings of a magnetic element, downdraft and upflows within the flux tube, and constant magnetic flux area oscillations are signatures of such a mechanism. Alternatively, the fourth option can be found in ``sausage'' modes \citep{1983SoPh...88..179E} excited through compression by granules. The very large amplitude of the oscillation in basically all variables, in particular in the magnetic field strength may also be a sign that it is not a true oscillation at all, but rather multiple episodes of convective collapse, with a loss of equilibrium in between. What is causing the flux tube's field to be so unstable has not become entirely clear from this study. However, the absence of a strong upflow during the decay phase of the field strength suggests that this phase is not initiated by a strong upflow (rebound shock) such as that found by \citet{1998A&A...337..928G}. Since we are only looking at a photospheric height, the present observations do not allow us to distinguish between the different mechanisms. In order to study the chromospheric response, we plan to supplement IMaX observations with the simultaneous Ca \textsc{ii} H filtergrams from the \textsc{Sunrise} Filter Imager \citep[SuFI;][]{2011SoPh..268...35G}.

It is evident that the formation and the subsequent evolution of a solar magnetic element is a complicated problem, where many phenomena take place, namely, emergence of a magnetic $\Omega$ loop, expulsion of its footpoints from a granule, merging of flux patches in a long-lived inflow, formation of a kG magnetic element by convective collapse and granular compression, a subsequent weakening of the field strength to further increase again to kG values either through an oscillation or a second collapse. For the first time, we have been able to observe and relate all these phenomena in a single example.

\acknowledgments

The work by I.S.R. has been funded by the Basque Government under a grant from Programa de Formaci\'{o}n de Personal Investigador del Departamento de Educaci\'{o}n, Universidades e Investigaci\'{o}n.
This work has been partially funded by the Spanish Ministerio de Econom\'{\i}a y Competitividad, 
through Project No. AYA2012-39636-C06, including a percentage from European FEDER funds.
The German contribution has been funded by the Bundesministerium f\"ur Wirtschaft und Technologie through Deutsches Zentrum f\"ur Luft- und Raumfahrt e.V. (DLR), grant number 50 OU 0401, and by the Innovationsfond of the President of the Max Planck Society (MPG). 
This work was partly supported by the BK21 plus program through the National Research Foundation (NRF) funded by the Ministry of Education of Korea.

\clearpage



\clearpage





\end{document}